# Classification of malware based on file content and characteristics

Mouhammd Alkasassbeh , Samail Al-Daleen

# Abstract

In general, the industry of malware has come to be a market which brings on loads of money by investing and implementing high end technology to escape traditional detection while vendors of anti-malware spend thousands if not millions of dollars to stop the malware breach since it not only causes financial losses but also emotional ones. This paper study the classification of malware based on file content and characteristics, this was done through use of Clamp Integrated dataset that includes 5210 instances. There are different algorithms were applied using Weka software, which are; ZeroR, bayesNet, SMO, KNN, J48, as well as Random Forest. The obtained results showed that Random Forest that achieved the highest overall accuracy of (99.0979%). This means that Random Forest algorithm is efficient to be used in malware classification based on file content and characteristics.

**Keywords; Malware, File content, ZeroR, bayesNet, SMO, KNN, J48, Random Forest**

## 1. Introduction

A malicious code which enters a computer without prior authorization is known as malware and it comes in various forms such as a virus, worm or simply a root kit, and since various schemes which are used to detect malware contain one, two, twenty and twenty eight signature based forms to identify malware, they are not able to detect those which are new or are variants of that malware. To discover a malware which is unknown, one is ought to analyze it in order to derive a signature; as if this signature were to become public, a lot of damage would have already occurred.

Recently, the industry of malware has come to be a market which brings on loads of money by investing and implementing high end technology to escape traditional detection while vendors of anti-malware spend thousands if not millions of dollars to stop the malware breach since it not only causes financial losses but also emotional ones .

One of the most difficult things to battle with today is the large number of data which needs to be scanned for malicious malware. An example of this is the anti-malware software provided by Microsoft which is existent in 160 million computers globally to scan 700 million devices per month. All this would create loads of data daily which needs to be scanned for malicious files; the reason as to why various volumes of the files is present is due to the polymorphism component which is introduced to avoid detection; meaning that the identical malware would be constantly modified and changed in order to make it harder to detect by making them similar to other files.

To analyze numerous files effectively and efficiently, these files need to be combined into groups to discover the family they belong to and then to apply the resulting criteria to all the files which are encountered on the devices to help discover those various malicious malware formats.

# 2. Related Work

In our previous work, we use machine learning for detection the abnormality of the network like in [17][18] [19][20]. Silvio Cesare and Yang Xiang [1] has suggested a control flow which can be used to identify various variants to provide a malware classification based on a flow graph through a static analysis which helps in the classification process. The used dataset includes 1000 malicious and 1000 benign, and SVR algorithm is applied. It is ineffective when the malware's code packing would transform to make it harder to be detected. In order to provide a demonstration on how affective an automated classification based on the flow graph and unpacking is, one is ought to implement it and evaluate it against real malware to result in a high level of accuracy in addition to the time for unpacking and revealing the codes hidden. The control flow and change detection also used for anomaly detection like in [16].

Liangboonprakong and Sornil [2] has discovered that identifying the malware family is rather a complex process which involves extracting specific characteristics out of malware samples, by presenting the n-gram features which were derived from the file contents, after that the patterns are specified and those of statistical nature would then get calculated and reduced using a method known as the sequential floating forward selection. Dataset in this paper includes 500 million files, and these files had been classified using KNN algorithm. Then, a classifier is implemented to help in determining the family of the malware to provide testing results which is evident that this method performs perfectly with an accuracy percentage of 96.64.

Bansal et al. [3] has observed that users usually get malware into their devices by either clicking on a malicious email link, by inputting a USF flash drive into their devices or by simply downloading a file from an infected website. In this paper, dataset includes 700 thousand files that had been addresses using SVR algorithm. Since the malware is increasing to a whopping rate, an investigation is ought to be conducted on smaller files in order to create a defense system of a larger scale proportions using malware classification of automated nature.

Ahmed and Lhee [4] has stated that there are various servers which do not allow executable codes in the network traffic of theirs, like blogger, YouTube, Picasa in addition to online shopping malls. As a result, those can be prevented from malware attacks or infections since there is observation to the ports to identify any packets which may be harmful. This research provides a scheme of the classification content to help identify such packets by analyzing the packet payload using two various steps. One is to analyze the packet payload to discover whether it contains data of multimedia type, where the used dataset includes 1000 files of malware. Also, the author used genetic algorithm to classify data. Then, classifying it either as an executable or perhaps a type of text; in this experiment, the scheme provided a very low rate of not only false positives of 2.53% but also of negative ones of a 4.69%. It should be noted that those inaccuracies would need investigation to be able to detect the malware presence, in addition to providing an analysis of combinational and statistical nature to explain those.

Santos [5] this research provides a scheme of the classification content to help identify such packets by analyzing the packet payload using two various steps. One is to analyze the packet payload to discover whether it contains data of multimedia type; Then, classifying it either as an executable or perhaps a type of text. Dataset in this paper includes 500 thousand files that were classified using KNN algorithm. In this experiment, the scheme provided a very low rate of not only false positives of 2.53% but also of negative ones of a 4.69%. It should be noted that those inaccuracies would need investigation to be able to detect the malware presence, in addition to providing an analysis of combinational and statistical nature to explain those.

Tabish et al. [6] has suggested a special procedure to help in discovering malware based on the byte-level depending on the content of the file. Therefore, the approach would be to compare it with the existing scheme to see if it will not memorizes the strings or byte-sequence which is existent in the file. Since this technique is not a signature based method then it would help in detecting the zero day or unknown malware and by operating this scheme using at the level of byte, it will not need any previous information regarding the file type. The author used dataset of 800 thousand files, and also used SVR and KNN algorithm to classify malware files. Putting this to the test by using a dataset of various files, six to be exact which are ZIP, DOC, PDF and the JPG in addition to the MP3 and EXE against 6 different malware types which are worms, viruses, backdoors and Trojans in addition to miscellaneous and constructor to provide a result which has demonstrated that this non-signature based has proven successful when compared with all the available techniques with a 90% accuracy rate.

Another researcher Ye et al. [7] has analyzed the contents of files which were taken from the sample files to use them to help discover malware, nut besides this it is known that "Downloader" is always accompanied by Trojans malware which provides very important details about this file sample. In this research, the file operations are studied and how they are able to provide malware detection results in order to further create a filing system using a classifier model to be known as a "Valkyrie". This is to combine the contents and relations of the file to help detect malware as this is considered the first attempt of its kind involving both of these properties of files. The used dataset in this paper obtained from "Comodo Cloud Security Center" that contains 37,930 user file, which shows file relations between samples of 30,950 malware, 225,830 of them are benign files and the rest of files are 434,870 unknown. Valkeryie system has provided with promising results of higher accuracy and effectiveness than any other software which are anti-malware like, VirusScan, AntiVirus and MacAfee in addition to Kaspersky and this system is currently being used as part of Comodo's Anti-Malware software.

Also Alabdulmohsin and HAN [8] stated that malware detection is studied by analyzing the characteristics or the relationships between the network distribution of files, and since this research has studied it at using a global malware delivery graph which combines the topology of the network in addition to the relationship of file dropping to suggest a Bayesian label propagation model which would combine these different information to include node features of content-agnostic nature. The authors used BFA algorithm to classify dataset of 1.2 million files. This approach would not require source code examination since it approximates the level of maliciousness within a file through a semi supervised procedure known as label propagation containing a the number of edges and nodes, and the linear time complexity. All this by looking at 567 million downloads to prove that this approach would provide high accuracy for malware detection.

McDaniel and Heydari [9] discussed three algorithms which are used to analyze the contents and type of files. In this paper, the dataset is got from the BFD. One is the ByteFrequency Analysis Algorithm (BFA) which calculates the distribution frequency to create a fingerprint specific to each file using the byte-frequency distribution, in addition to calculating the strength correlation since it is considered a factor in this process. Using these three algorithms to compare the files and the fingerprints created to discover the file type as it reported an accuracy level of 27.50 for BFA, 45.83 for BFC and lastly, 95.83% for FHT algorithm.

Karampatziakis et al. [10] provided a presentation a classification system for detecting malware depending on the graph existent in the file relationships and to provide authentication for the concept, contaminated relationships are analyzed to provide extensive data. The paper has 3.4 million files that are included in a dataset, which was analyzed. The methodology would be general and depends on the estimate initially regarding various files existent in the data in addition to the information propagating at the sides of the graph by the use of the Max neighbor algorithm; meaning that it can also be implemented to various other file relationships. A clear example of this would be, the false negative rate was lowered from 42.1% to 15.2% keeping the false positive rate at a 0.02%. This system would be considered as scalable, as the application can provide great classifiers out of a big graph that contains more than 719 thousand containers in addition to 3.4 million files all in the small time of 16 minutes.

Chen et al. [11] Based on the idea that the contents of the file derived from its sample, in the binary strings, instruction sequences, API, known as Application Programming Interface, and data mining methods like the Vector Machines and Naive Bayes as it was implemented to detect malware. Here, a study regarding the way file relations are implemented to enhance the rate of detection in results in addition to creating a system of verdict known as "Valkyrie". This would be created on classifier model that is semi-parametric in order to add the file relations and its combination to one another in order to detect malware. This is by far the first work known, which would use both the file relation and detection to discover malware. An experimental study regarding a huge set of PR files that were extracted from antimalware products of the company of Comodo Security Solutions as it aims to compare many approached for detecting malware, where the dataset was includes 30000 files. This resulted in providing efficiency and accuracy regarding the Valkyrie system as it is better when compared to the famous anti-malware software like McAfee VirusScan and Kaspersky AntiVirus, in addition to various detection systems that are data mining based.

In [12], Kolter et al applied data mining and n-gram approaches in order to discover malicious executables existent in the wild. The n-gram analysis was implemented to take out the features out of 1, 651 malicious and 1, 971 benign PE files. Such files were gathered out of machine running Windows XP and 2000 operating systems. Such PE files that are malicious were extracted from the VX Heavens Virus Collection's older versions. The accuracy of detection was reported as the area existent below the ROC curve (AUC) that is a measurement that is completed when compared to the accuracy of detection. The AUC had demonstrated that the decision trees boosted would outperform the remaining classifiers for problems of classification.

Stolfo et al, in the seminal work implemented the analysis of n-gram regarding the identification of file type [13] to be used for malware detection later. In previous work that fingerprints analysis, implementing the 1-gram byte distribution regarding the file as a whole and then put it to comparison with the various file type in order to determine its type; The distribution of the 2-gram and 1-gram in order to experiment it in a dataset that contains 31 benign application executable, where there were 331 benign executables and 571 viruses, within the System32 folder. The resulted concluded that the suggested method would detect an appropriate segment of the malicious files. One must note that this method is limited to the embedded malware and would not relate to the stand-alone malware detection.

In [14], Schultz et al applied various methods of data mining in order to separate the malicious and benign executables in the MS-DOS or Windows format existent in the dataset containing 3, 265 malicious executables and 1, 001 benign. Such executables contained samples of the portable executables file of PE format containing 38 malicious and 206 benign. An algorithm that is inductive and rule learning named as RIPPER that is implemented in various feature vectors implemented for classification. Such schemes would be dependent on the information of DLL to give an accuracy of detection with an 89.07%, 83.62%, 88.36% as follows.

# 2. Dataset &Experiments Results
## 2.1 Dataset Descrbtion:

As previously mentioned, this paper aims to study Classification of malware based on file content and characteristics. So, ClaMP_Integrated dataset had been used [15], where this dataset includes 5210 instances. Table (1) represents a description for the used dataset in this paper.

Also, as previously mentioned, the total samples was 5210, where (2722) of them were malware and the rest of files of (2488) were Benign. Moreover, the total Features was (69), where (54) of them were Raw Features as well as (15) of them were Derived Features.

| Name of datasets | ClaMP_Integrated | Minimum | 0 |
|---|---|---|---|
| Number of instances | 5210 | Maximum | 37008 |
| Attributes | 70 | Mean | 152.659 |
| Sum of weights | 5210 | StdDev | 616.499 |
| Distinct | 10 | Unique | 3(0%) |

Tabel (1): Description Dataset

In order to classify files of dataset if there is malware contents or no based on file content and characteristics, there are different algorithm were used, which are; ZeroR, bayesNet, SMO, KNN, J48, as well as Random Forest.

## 2.2 Results and Discussion

The entire previously mentioned algorithm had been applied into Weka software in order to know their performance and accuracy in malware classification, as illustrated in Table (2).

| Algorithm | Precision | FP Rate | TP Rate | Overall Accuracy | PRC area | Roc Area | F-measure | Recall | Overall Accuracy |
|---|---|---|---|---|---|---|---|---|---|
| ZeroR | 0.273 | 0.522 | 0.522 | **52.2457%** | 0.501 | 0.5 | 0.359 | 0.522 | **52.2457%** |
| bayesNet | 0.947 | 0.055 | 0.946 | **94.6449%** | 0.985 | 0.985 | 0.946 | 0.946 | **94.6449%** |
| SMO | 0.957 | 0.044 | 0.956 | **95.643%** | 0.936 | 0.956 | 0.956 | 0.956 | **95.643%** |
| KNN | 0.979 | 0.022 | 0.979 | **97.8503%** | 0.968 | 0.979 | 0.979 | 0.979 | **97.8503%** |
| J48 | 0.979 | 0.022 | 0.979 | **97.8695%** | 0.972 | 0.982 | 0.979 | 0.979 | **97.8695%** |
| Random Forest | 0.991 | 0.009 | 0.991 | **99.0979%** | 0.999 | 0.999 | 0.991 | 0.991 | **99.0979%** |

Table 2 represents results of implementation for all algorithms and their overall

. Table (2) represents results of implementation for all algorithms and their overall accuracy in malware classification. As shown in Table 2, lowest accuracy was for implementation of ZeroR algorithm with overall accuracy of (52.2457%) that considered low when it compared with accuracy of other algorithm especially Random Forest that achieved the highest overall accuracy of (99.0979%). This means that Random Forest algorithm is efficient to be used in malware classification based on file content.

**Figure (1):Comparison between algorithms**

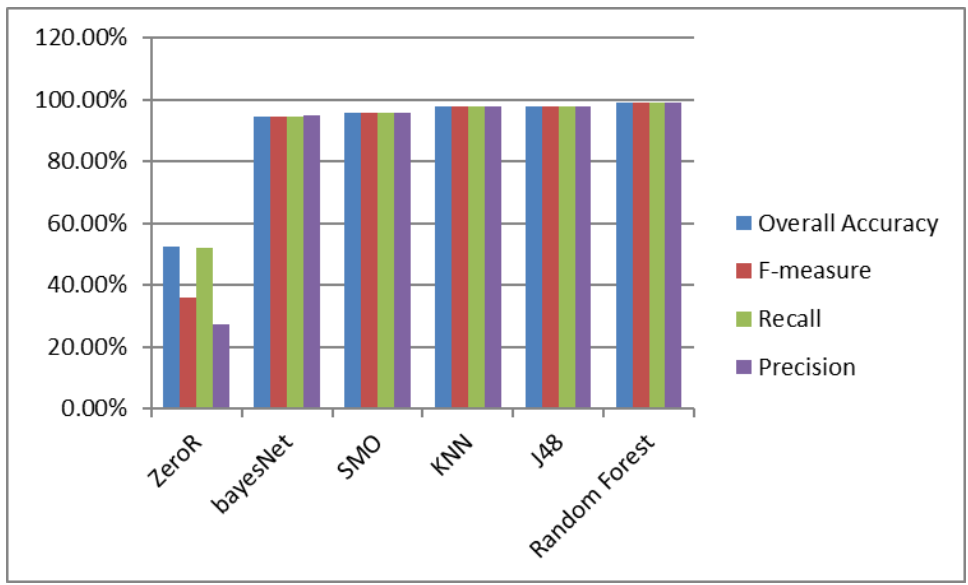

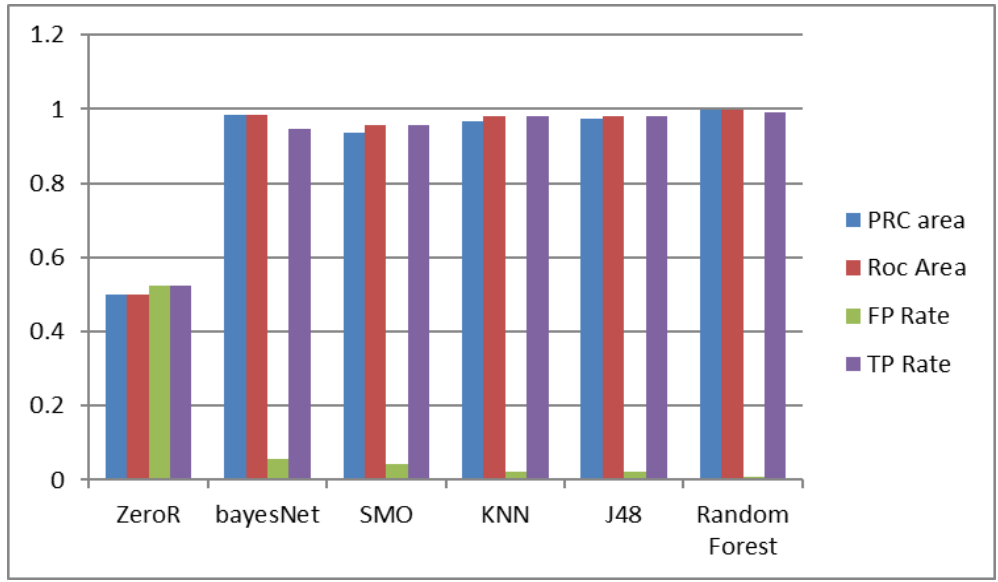

*Figure (1):Comparison between algorithms (other features)*

Figures (1) and (2) represent a comparison between all algorithms in detail that gathered to obtain the overall accuracy of algorithm. As shown in the figures, it is clear that false positive ratio (FP ratio) of Random Forest algorithm, which achieved the best accuracy, is very low referring to the number of negative files that categorized wrongly as positive is low. Regarding to precision-recall curves (PRC) represented in Table 2 and Figure 2, value of PRC for Random Forest algorithm is (0.999) referring high performance of algorithm in general, and more useful than Receiver Operating Characteristics (ROC) of (0.999) that also evaluate performance of Random Forest algorithm. Regarding to True Positive (TP) of (0.991) that referring to 99% of files are correctly identified. Also, Table 2 and Figure 1 show that precision value of (0.991) referring to positive predictive files using Random Forest algorithm. Sensitivity of algorithm was clear through value of recall that equal (0.991). Finally, the high accuracy that defined as the weighted harmonic mean of the precision and recall (F-Measure) was (0.991).

# 3.Conclusion

To discover a malware which is unknown, one is ought to analyze it in order to derive a signature; as if this signature were to become public, a lot of damage would have already occurred. One must also discover the malware on its 0-day because effectively by using a procedure which compares characteristics of various malware by using the executable codes. One of the most difficult things to battle with today is the large number of data which needs to be scanned for malicious malware.

Results of implementation showed that lowest accuracy was for implementation of ZeroR algorithm with overall accuracy of (52.2457%) that considered low when it compared with accuracy of other algorithm especially Random Forest that achieved the highest overall accuracy of (99.0979%). This means that Random Forest algorithm is efficient to be used in malware classification based on file content and characteristics.

# 4. Recommendations

Based on the obtained results, there are many recommendations to be considered for future works, as the following:

- To apply other algorithms such as genetic algorithms in order to achieve higher accuracy.
- To use Matlab program for implementation and apply the algorithm in large dataset to show its performance in case of huge datasets.